\documentclass[twocolumn,pra,a4paper,nofootinbib,showpacs,aps,floatfix]{revtex4}
\usepackage{graphicx}
\usepackage{bm}
\usepackage{amsmath}
\def\la{\langle}
\def\ra{\rangle}

\begin{document}
\title{\bf{Avoided crossing resonances: structural and dynamical aspects}}
\author{I. Lizuain}
\email[Email address: ]{ion.lizuain@ehu.es}
\affiliation{Departamento de Qu\'\i mica-F\'\i sica,
Universidad del Pa\'\i s Vasco, Apdo. 644, Bilbao, Spain}
\author{E. Hern\'andez-Concepci\'on}
\email[Email address: ]{ethelhc@ull.es}
\affiliation{Departamento de F\'\i sica Fundamental II, 
Facultad de F\'\i sica, Universidad de La Laguna}
\author{J. G. Muga}
\email[Email address: ]{jg.muga@ehu.es} 
\affiliation{Departamento de Qu\'\i mica-F\'\i sica,
UPV-EHU, Apdo. 644, Bilbao, Spain} 
\begin{abstract}

We examine structural and dynamical properties of quantum resonances associated with an avoided crossing and identify the parameter shifts where these properties 
attain extreme values.
Thus the concept of avoided crossing resonance can be defined in different ways, which do not coincide in the general case.
These definitions are described first at a general level, and then 
for a two-level system coupled to a harmonic oscillator, of the type 
commonly found in quantum optics. 
Finally the results obtained are exemplified and applied to optimize the fidelity and speed of quantum gates in trapped ions. 

\end{abstract}
\maketitle

\section{Introduction}
%

In this paper we study the definition of a resonance associated with an avoided crossing. 
Avoided crossings are quite common in many fields of nuclear, atomic, or molecular physics such as 
laser driven trapped ions \cite{CBZ94,LM07}, two level atoms coupled to a cavity mode \cite{cohen73}, or diamagnetic hydrogen in magnetic fields \cite{WDW98}.
In the avoided crossing regions,
two eigenvalues of the system approach first as a 
parameter $\xi$ is varied, but then veer from each other. The ``bare levels'' of a
zeroth order Hamiltonian 
do cross at a reference value $\xi_0$, but in the full Hamiltonian a perturbation 
connecting them causes the splitting.   
The eigenvalues also 
interchange their character along the avoided crossing: each of the two eigenvalues is dominated by different bare levels 
before and after the crossing. The resonance is also characterized by maximal  
transition probabilities among the bare levels.
As we shall see, however, the central loci of these phenomena do not generically  
coincide, so 
different ``shifts'' will be introduced. The shifts are not only due to the two bare levels directly implied, 
but to the ``contamination'' or influence of the rest of the levels.   
  
The article is organized in three increasing levels of concreteness: we shall
first discuss formal general aspects in Sec. \ref{general_theroy_section}, then make a more specific analysis
for Hamiltonians commonly found in quantum optics which descibe a two-level system
coupled to a harmonic oscillator in Sec. \ref{2_level_harmonic_section}, and finally exemplify and apply the results by 
optimizing the fidelity and speed of quantum gates for trapped ions in Sec. \ref{ss_tuning_sec}. The paper ends with Sec. \ref{ss_gate_discussion}, a brief discussion of the results obtained, and a technical Appendix.

%
\section{Generic and formal aspects}
\label{general_theroy_section}
\begin{figure}[t]
\begin{center}
\includegraphics[width=8cm]{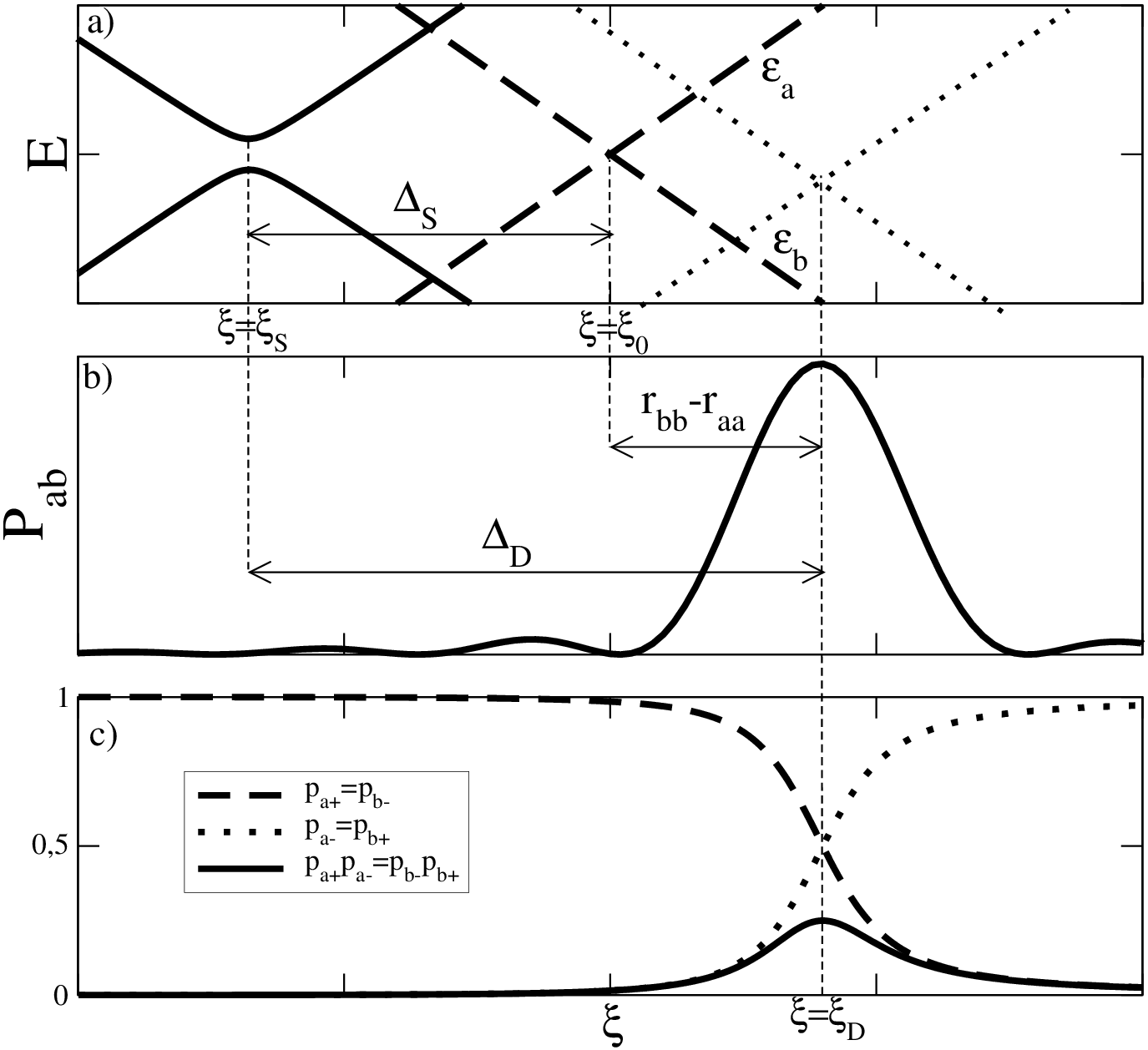}
\caption{(a) Schematic plot of the energy levels around the resonance between the $|a\ra$ and $|b\ra$ bare states. 
Non-perturbed (dashed lines) energy levels $\epsilon_a$ and $\epsilon_b$ cross each other at the bare resonance $\xi=\xi_0$.
The perturbation $V$ shifts the position of this resonance (structural definition)
to $\xi=\xi_{S}$ (solid lines).
(b) According to a dynamical definition the resonance is instead at $\xi=\xi_{D}$, where the state-flip probability is maximum.
The dynamical shift $\Delta_D$ is the separation between both definitions, i. e., $\Delta_D=\xi_D-\xi_S$. (c) The change of character of the dressed energy levels occurs at $\xi_D$. }
\label{resonances_general_fig_abc}
\end{center}
\end{figure}

Consider the general Hamiltonian
\begin{equation}
H=H_0(\xi)+V(\xi),
\end{equation}
where $V(\xi)$ is a small perturbation of the non-perturbed (bare) Hamiltonian $H_0$. Both parts may depend on an external parameter $\xi$ and the eigenenergies and eigenstates of $H_0$ (bare energies and states) are supposed to be known,
\begin{equation}
 H_0|\alpha\ra=\epsilon_\alpha|\alpha\ra.
\end{equation}
with $\epsilon_\alpha=\hbar\omega_\alpha$. Assume also that the energy levels corresponding to two given bare states $|a\ra$ and $|b\ra$ cross each other at $(\xi_0,E_0)$ in the $(\xi,E)$ plane (Fig. \ref{resonances_general_fig_abc}a, dashed lines). This defines the bare or unperturbed resonance at $\xi=\xi_0$.
If both states are  connected by the perturbation, $V_{ab}=\la a|V|b\ra\neq0$, the crossing between these levels will become an avoided-crossing. 

%
%
\subsection{Structural Shift}
Let us now assume that the two energy levels $\epsilon_{a}$ and $\epsilon_{b}$ are close to each other but far from other levels, i. e., a well isolated avoided crossing. The perturbed energy levels may be \emph{exactly} given by an 
effective Hamiltonian \cite{cohen73,CDG98,LME08} in the subspace spanned by the states $|a\ra$ and $|b\ra$ given by
\begin{eqnarray}
H_{eff}&=&\left(  \begin{array}{cc}
\epsilon_a+R_{aa} & R_{ab}\\
R_{ba} & \epsilon_b+R_{bb}
              \end{array}\right)\nonumber\\
&=&
\hbar\left(  \begin{array}{cc}
-\delta & r_{ab}\\
r_{ba} & \delta
              \end{array}\right),
\label{Heff}
\end{eqnarray}
where the effective detuning $\delta$ is defined as
\begin{eqnarray} 
\label{effective_det_gral}
\delta&\equiv&\frac{1}{2}\left(\omega_{b}-\omega_a+r_{bb}-r_{aa}\right)
\end{eqnarray}
and $r_{\alpha\beta}$ are the matrix elements of the level shift operator
divided by $\hbar$, 
\begin{eqnarray}
\label{level_shift_operator_ssgatte}
r(E)&=&\frac{R(E)}{\hbar}\\
R(E)&=&PVP+\sum_{n=1}^{\infty}PV\left(\frac{Q}{E-H_0}V\right)^nP,
\end{eqnarray}
with $P=|a\ra\la a|+|b\ra\la b|$ and $Q=\mathbf{1}-P$. 
Even though it is 2-dimensional, the effective Hamiltonian (\ref{Heff}) contains information about the whole Hilbert space via the operator $Q$, the projector onto the non-resonant subspace. 
$H_{eff}$ is an implicit Hamiltonian, since it depends on the exact perturbed energy $E$ through $R(E)$.
A first approximation is to evaluate $H_{eff}$ at the nonperturbed position of the resonance, i. e., at $\xi=\xi_0$ and $E=E_0$. Further corrections can be obtained by iteration \cite{cohen73}.

The position of the resonance may be defined to be at the center of the anti-crossing \cite{cohen73,CDG98}, which may be found at the extrema of the dressed energy levels, i. e., at $\xi=\xi_{S}$ in Fig. \ref{resonances_general_fig_abc}a. 
This is also the point of minimal splitting. Using the expression of the eigenergies of $H_{eff}$, $\epsilon_\pm=\pm(\delta^2+|r_{ab}|^2)^{1/2}$, $\xi_{S}$ is found 
from the condition
\begin{equation}
 \frac{\partial}{\partial\xi}\left(\delta^2+|r_{ab}|^2\right)^{1/2}=0,
\end{equation}
or more explicitly
\begin{eqnarray}
\label{SR_condition}
2\delta\frac{\partial\delta}{\partial\xi}+\frac{\partial\left|r_{ab}\right|^2}{\partial\xi}=0, 
\end{eqnarray}
%
see Fig. \ref{resonances_general_fig_abc}a.

The discussion so far refers to an energy level structure criterion so that 
$\Delta_S\equiv\xi_S-\xi_0$ is a ``structural shift''. 
A different definition, based on the dynamics of the system, will be 
discussed in the following.  

Before, note that if $PVP$ does not contribute,
$\Delta_S$
is a generalization of the Bloch-Siegert (BS) shift, first described by Bloch and Siegert while studying the failure of the Rotating Wave Approximation  \cite{BS40,cohen73}, and due entirely to the effect of non-resonant transitions involving levels in the subspace $Q$. 
%
%
\subsection{Dynamical Shift}
Let us now prepare the system in the $|a\ra$ bare state and look for the probability to find it in $|b\ra$. We could define the resonance in a dynamical way as the value of $\xi$ where the state-flip probability is maximal.
For the dynamics governed by the effective Hamiltonian 
(\ref{Heff}),\footnote{The exact dynamics in the $P$-subspace is given by a non-Markovian 
equation with a memory kernel, and $H_{eff}$ provides the
Markovian approximation.} the state-flip probability is given by 
%
\begin{eqnarray}
P_{ab}&=&\left|\la b|e^{-iH_{eff}t}|a\ra\right|^2\nonumber\\
&=&\frac{|r_{ab}|^2}{\delta^2+|r_{ab}|^2}\sin^2 \frac{\Omega t}{2},
\end{eqnarray}
where $\Omega=(\delta^2+|r_{ab}|^2)^{1/2}$. If a $\pi$-pulse (defined by $\Omega t=\pi$) is applied, this probability shows a maximum for  $\delta=0$, at $\xi= \xi_{D}$ defined by the condition
%
\begin{equation}
\label{DR_condition}
 \delta=\omega_b-\omega_a+r_{bb}-r_{aa}=0,
\end{equation}
see Fig. \ref{resonances_general_fig_abc}b. 

The dynamical definition of the resonance may be also understood in terms of the  change of the character of each dressed energy level. Consider the eigenstates of $H_{eff}$
\begin{eqnarray}
|\epsilon_+\ra&=&\cos\frac{\theta}{2}e^{i\phi/2}|a\ra+  \sin\frac{\theta}{2}e^{-i\phi/2}|b\ra,\nonumber\\
 |\epsilon_-\ra&=&-\sin\frac{\theta}{2}e^{i\phi/2}|a\ra+  \cos\frac{\theta}{2}e^{-i\phi/2}|b\ra,
\end{eqnarray}
where $\tan\theta=-|r_{ab}|/\delta$ and $r_{ab}=|r_{ab}|e^{i\phi}$. 
The ``character'' of the dressed states around resonance is given by $p_{\alpha,\pm}=|\la \alpha|\epsilon_\pm\ra|^2$, the projection of the bare state $|\alpha\ra$($\alpha=a,b$) onto the dressed state $|\epsilon_\pm\ra$. Then we have that
\begin{eqnarray}
p_{a+}&=&p_{b-}=\cos^2\frac{\theta}{2},\nonumber\\
p_{a-}&=&p_{b+}=\sin^2\frac{\theta}{2}.
\end{eqnarray}
The character change of the dressed levels is centered at $\theta=\pi/2$,  when both projections are equal, $p_{\alpha +}= p_{\alpha-}$. This corresponds to $\delta=0$, and coincides with the dynamical resonance condition (\ref{DR_condition}), see Fig. \ref{resonances_general_fig_abc}c. $\xi_{D}$ may thus be defined  independently  of the $\pi$-pulse condition. Numerical calculations diagonalizing full $n$-dimensional Hamiltonians (with $n$ large enough to assure convergence) 
confirm this result,  
and examples are provided in Sec. \ref{ss_tuning_sec} below. 

To summarize the main results obtained so far: the resonance location may be defined in different ways: a ``structural'' criterion, Eq.  (\ref{SR_condition}), gives $\xi_{S}$, where the splitting between the dressed energy levels is minimal,
whereas the ``dynamical" criterion gives  $\xi_{D}$, defined by the condition (\ref{DR_condition}), where the state-flip probability is maximal.
$\xi_S$ and $\xi_D$ do not coincide in the general case, since the two conditions cannot be simultaneosly fullfilled as long as $|r_{ab}|$ depends on $\xi$. 
The two values are separated by the ``dynamical shift'' $\Delta_D\equiv\xi_D-\xi_S$, proportional to $\partial |r_{ab}|^2/\partial\xi$.  

\section{Two-level system coupled to a harmonic oscillator}
\label{2_level_harmonic_section}
One of the simplest cases in which the difference between the structural and dynamical definitions of a resonance can be observed is a two level system coupled to a harmonic oscillator and driven by some external field, as frequently found in quantum optics.  
The Hamiltonian, in an appropriate interaction picture takes the form
\begin{equation}
\label{2_level_harmonic_ham}
H=\hbar\omega a^\dag a+\frac{\hbar\xi}{2}\sigma_z+V(\xi),
\end{equation}
where $V(\xi)$ is assumed to be a small perturbation of the unperturbed Hamiltonian $H_0=\hbar\omega a^\dag a+\hbar\xi\sigma_z/2$. 
$a^\dag$ and $a$ are the usual creation and annihilation operators for the harmonic oscillator, and the Pauli atomic inversion operator is 
$\sigma_z=|a\ra\la a|-|b\ra\la b|$. 
$V$ may also depend on $\sigma_i$ with $i=x,y,z$.
The bare energy levels of this system are   
\begin{eqnarray}
\epsilon_{a,n}&=&n\hbar\omega+\frac{\hbar\xi}{2},\nonumber\\
\epsilon_{b,n}&=&n\hbar\omega-\frac{\hbar\xi}{2},
\end{eqnarray}
$n$ being the vibrational quantum number $n=0,1,2\hdots$. The energy levels corresponding to two given bare states $|a,n_a\ra$ and $|b,n_b\ra$
cross each other at 
\begin{eqnarray}
\xi_0&=&(n_b-n_a)\omega,\\
E_0&=&(n_a+n_b)\frac{\hbar\omega}{2},
\end{eqnarray}
see Fig. \ref{resonances_general_fig_abc}a, dashed lines. 
As pointed out in the previous section, the perturbation will not only split the energy levels but it will also shift the position of the anti-crossing from $\xi_0$ to $\xi_S$, see Fig. \ref{resonances_general_fig_abc}a, solid lines. 
It may also happen that both levels are shifted ($r_{aa},r_{bb}\neq0$) but not splitted ($r_{ab}=0$), and thus the crossing remains permitted, as in Fig. \ref{resonances_general_fig_abc}a, dotted lines.
The perturbed energy levels will be described by the effective Hamiltonian (\ref{Heff}), with the effective detuning (\ref{effective_det_gral}),
\begin{eqnarray} 
\label{effective_det_2level_harmonic}
\delta=\frac{1}{2}\left(\xi_0-\xi+r_{bb}-r_{aa}\right).
\end{eqnarray}
The position of the resonance according to structural and dynamical 
criteria will be determined 
by the conditions (\ref{SR_condition}) and (\ref{DR_condition}), 
\begin{eqnarray}
\label{SR_2level_harmonic}
\xi_{S}&=&\xi_0+r_{bb}-r_{aa}+\Delta_D(\xi),\\
\label{DR_2level_harmonic}
\xi_{D}&=&\xi_0+r_{bb}-r_{aa},\\
\Delta_D(\xi)&=&\frac{2\frac{\partial\left|r_{ab}\right|^2}{\partial\xi}}{\frac{\partial r_{bb}}{\partial\xi}+\frac{\partial r_{aa}}{\partial\xi}-1}.
\label{2_level_dyn_shift}
\end{eqnarray}
where, as a first approximation, all the matrix elements of $r$
and their derivatives are evaluated at $\xi_0$. 

\section{Tuning Quantum Gates for Maximal Speed}
\label{ss_tuning_sec}
The efficient physical implementation of quantum gates and quantum information
processing is a major goal for different fields of physics. The 
approaches based on ions in a linear trap pioneered by 
Cirac and Zoller \cite{CZ95} are among the most developed, and 
have become a working horse to test basic
quantum information processing \cite{kaler03}.
In addition, the formalism is very
similar or even equal in some limits 
to the one applied in other systems, such as cavity QED \cite{Solano},
or superconducting qubits \cite{Deppe08}.

Quantum gates based on trapped ions illuminated by lasers can be speeded up considerably by adjusting the laser to the exact position of the resonance \cite{steane00}. In these quantum gates one is interested in obtaining the maximum
fidelity, that is, the maximum transition probability from one state to another.
The dynamical shift defined above plays then an important role, since 
the laser parameters have to be adjusted to $\xi_{D}$ and not to $\xi_{S}$, unless of course they coincide.  

Let us now consider an effectively 1D trapped ion interacting with a classical field in a laser adapted interaction picture and after applying the optical Rotating Wave Approximation (RWA). It is described by the
Hamiltonian
\begin{equation}
\label{full_ham}
H=\hbar\omega_T a^\dag a -\frac{\hbar\Delta}{2}\sigma_z+\frac{\hbar\Omega_R}{2}\left[e^{i\eta(a+a^\dag)}\sigma_++H.c\right],
\end{equation}
where $\Delta=\omega_L-\omega_0$ is the detuning 
(laser frequency minus transition frequency between levels $|g\ra$ and $|e\ra$),
$\eta=(\omega_R/\omega_T)^{1/2}$ is the Lamb-Dicke (LD) parameter, 
and $\sigma_z=|e\ra\la e|-|g\ra\la g|$;  
$\omega_R$ is the recoil frequency of the ion and $\Omega_R$ is assumed 
real without loss of generality. 
%
%
\subsection{Stark Shift gate}
The dynamical shift is clearly observed in the so-called ``Stark shift gate'', proposed by Jonathan, Plenio and Knight \cite{JPK00}. 
Looking for faster quantum gates, they proposed a scheme where high intensity lasers overcome the slowness problem of the Cirac-Zoller (CZ) \cite{CZ95} and related gates, where low intensity lasers (in the sense that $\Omega_R\ll\omega_T$) limit the gate velocity. 
Setting the laser frequency on resonance ($\Delta=0$) and the laser intensity so that the Rabi frequency $\Omega_R$ and the trapping frequency $\omega_T$ coincide (first ``Rabi Resonance'' \cite{LM07}, $\Omega_R=\omega_T$), the splitting of the dressed states coincides with 
one vibrational
quantum, so we can expect avoided level crossings and fast and efficient vibronic transitions.    
This gate is thus based on a double resonance  condition for both laser frequency and
intensity.

The Hamiltonian given in Eq. (\ref{full_ham}) is written in the basis of the eigenstates of $\sigma_z$ $\lbrace|g\ra,|e\ra\rbrace$ which form the computational basis in the CZ gate scheme. The Stark-Shift gate, on the other hand, works in the $\lbrace|\pm\ra\rbrace$ basis, where $|\pm\ra=(|g\ra\pm|e\ra)/\sqrt2$ are the eigenstates of $\sigma_x$. It is therefore convenient to write $H$ in this new basis,
\begin{eqnarray}
\label{SS_gate_ham}
 \tilde H&=&\hbar\omega_T a^\dag a +\frac{\hbar\Delta}{2}\tilde\sigma_x\nonumber\\
&+&\frac{\hbar\Omega_R}{2}\left[\cos{\eta(a+a^\dag)}\tilde\sigma_z+i\sin{\eta(a+a^\dag)}(\tilde\sigma_+-\tilde\sigma_-)\right],\nonumber
\end{eqnarray}
where we have redefined the Pauli operators in the $\lbrace|\pm\ra\rbrace$ basis according to
$\tilde\sigma_+=|+\ra\la-|=\tilde\sigma_-^\dag,\tilde\sigma_x=\tilde\sigma_++\tilde\sigma_-,\tilde\sigma_z
=|+\ra \la +| - |-\ra \la -|$. These tilde operators are related to the usual Pauli operator by the transformation
\begin{eqnarray}
\sigma_z&=&-\tilde\sigma_x,\\
\sigma_+&=&\frac{1}{2}\left(\tilde\sigma_z+\tilde\sigma_+-\tilde\sigma_-\right)=\sigma_-^\dag,\\
\sigma_x&=&\tilde\sigma_z.
\end{eqnarray}
Both Hamiltonians $H$ and $\tilde H$ are completely equivalent, only a change of basis has been applied. In order to allow for lasers with arbitrary intensity  we consider the LD parameter to be the perturbative parameter. Then,  we may split $\tilde H$ as
$H_0+V(\eta)$ with   
\begin{eqnarray}
H_0&=&\hbar\omega_Ta^\dag a+\frac{\hbar\Omega_R}{2}\tilde\sigma_z,\nonumber\\
V(\eta)&=&\frac{\hbar\Omega_R}{2}\left[(\cos\hat\alpha-1\right)\tilde\sigma_z+
i\sin\hat\alpha\left(\tilde\sigma_+-\tilde\sigma_-\right)],
%
\label{SS_ham_splitted}
\end{eqnarray}
where $\hat\alpha\equiv\eta(a+a^\dag)$  and where $\Delta=0$ as required by the SS-gate (the perturbation vanishes in the LD limit, $V(\eta=0)=0$).  This SS-gate Hamiltonian has the same form of the general Hamiltonian (\ref{2_level_harmonic_ham}), so the formalism described in Sec. \ref{2_level_harmonic_section} may be applied by redefining $|a\ra=|+,n\ra$, $|b\ra=|-,n+1\ra$ and $\xi=\Omega_R$.

%
\begin{figure}[t]
\begin{center}
\includegraphics[width=7cm]{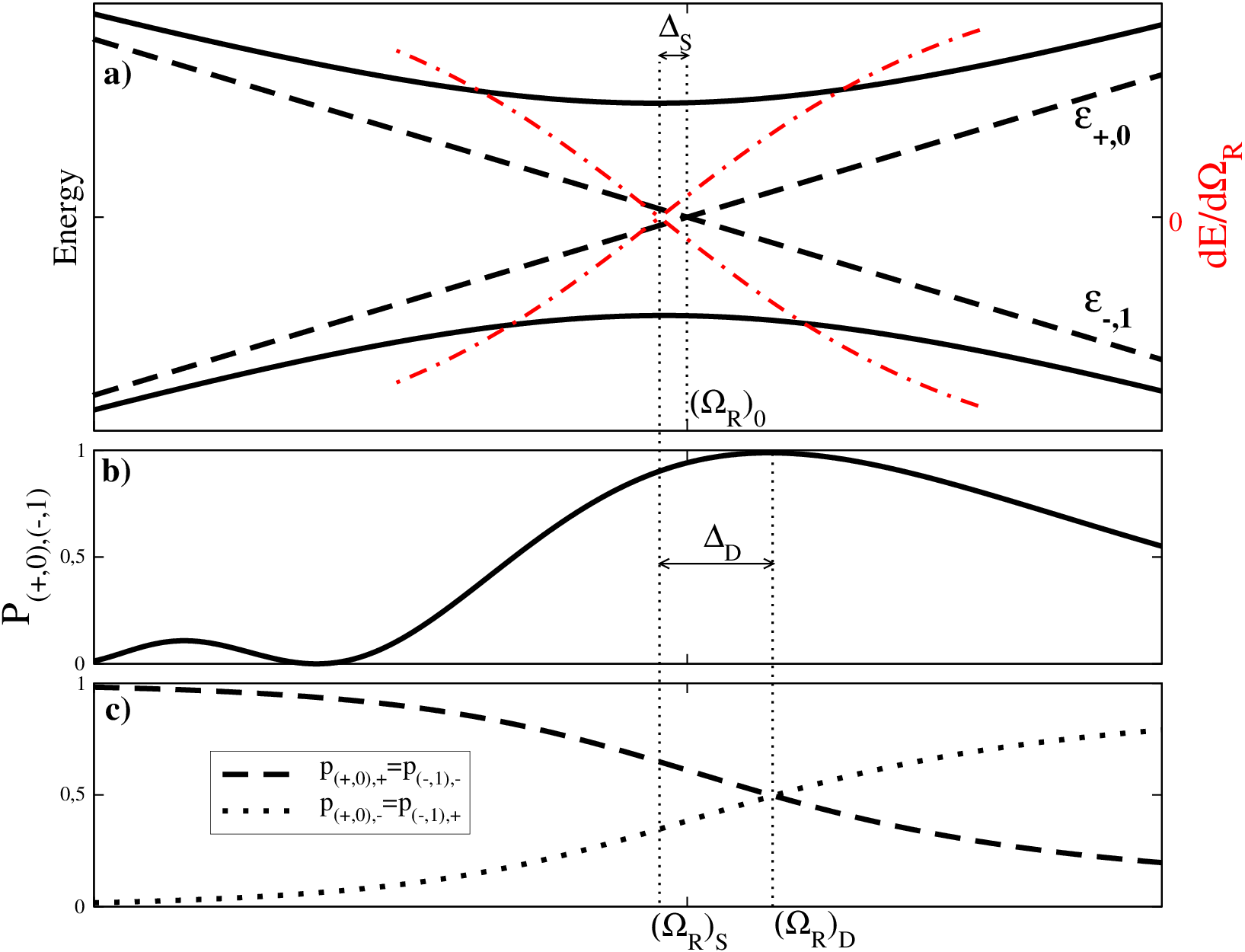}
\caption{(Color online) Stark Shift gate scheme: (a) Energy levels of a trapped ion coupled to a resonant ($\Delta=0$) laser around the first Rabi Resonance where the Stark Shift gate transition  $|+,0\ra\leftrightarrow|-,1\ra$ takes place.
The non perturbed or bare ($\eta=0.0$) energy levels $\epsilon_{+,0}$ and $\epsilon_{-,1}$ cross each other at $(\Omega_R)_0$, dashed lines. The perturbed energy levels ($\eta=0.3$) form an avoided crossing centered at $(\Omega_R)_S$, and shifted from $(\Omega_R)_0$ by $\Delta_S$, solid lines. The dash-dotted (red) lines represent the derivative of the perturbed energy levels with respect to $\Omega_R$. They cross each other at $(\Omega_R)_S$, where the derivative vanishes (zero slope).
(b) When a $\pi$-pulse (defined as $\eta\Omega_Rt=\pi$) is applied, the state-flip probability from $|+,0\ra$ to $|-,1\ra$ shows a maximum at $(\Omega_R)_D$, shifted from $(\Omega_R)_S$ by the dynamical shift $\Delta_D$.
(c) The change of character of the dressed energy levels occurs at $(\Omega_R)_D$.
All calculations have been done by numerically diagonalizing the full Hamiltonian (\ref{full_ham}) to all orders in $\eta$ and including a large number of vibrational states ($n=25$).}
\label{resonances_ss_fig}
\end{center}
\end{figure}
\subsection{The stark-shift gate error and velocity}
\begin{figure}[t]
\begin{center}
\includegraphics[width=7cm]{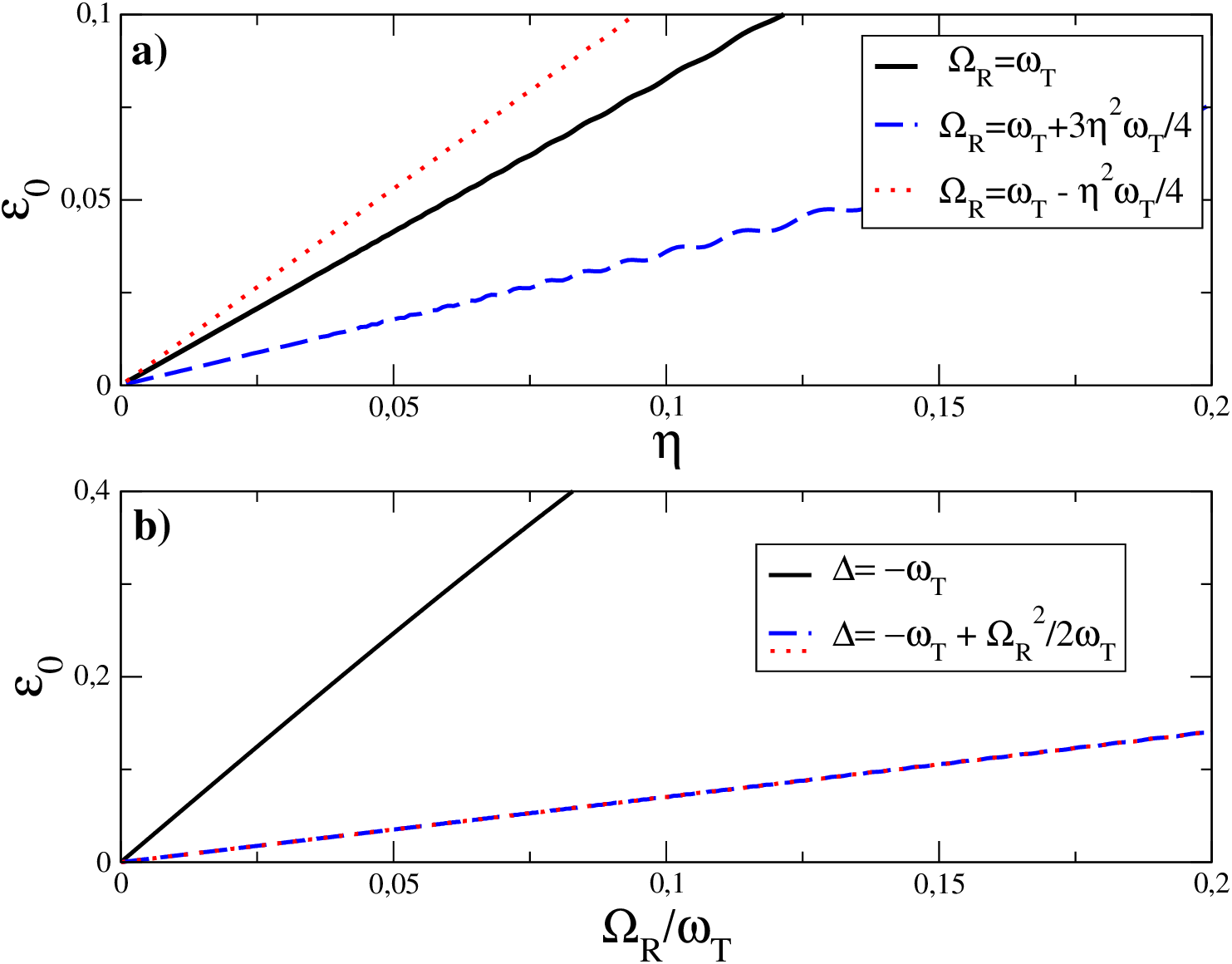}
\caption{(Color online) Error of the quantum gate after a $\pi$-pulse 
as the perturbation becomes stronger
(a) in the SS-gate scheme and (b) in the CZ scheme with $\eta=0.1$. 
Solid (black) lines: the laser is tuned to the bare resonance without any correction. 
Dotted (red): the laser is tuned to the structural resonance.
Dashed (blue): tuning the laser to the structural resonance in the SS-gate gives a larger error than the uncorrected gate, this shows the importance of the ``dynamical" shift. In the CZ scheme (b) the dynamical shift is negligible so the dotted (red) and dashed (blue) lines coincide.
In both gates, the dynamics with the effective Hamiltonian   
would give a zero error at the dynamical resonance. The
discrepancy with the exact calculation is due to the 
non-Markovian character of the true evolution.}
\label{error_eta_fig}
\end{center}
\end{figure}

Let us now consider the Stark-Shift gate interaction between computational states $|+,n\ra$ and $|-,n+1\ra$.
If the system is prepared initially in the state $|+,n\ra$, an ideal Stark-Shift operation would flip the state of the system to
$|-,n+1\ra$. We define the gate error as the imprecision \cite{steane00} given by 
\begin{equation}
\epsilon_n=\left[1-P_{|+,n\ra\rightarrow|-,n+1\ra}\right]^{1/2}. 
\end{equation}
Numerical simulations with a large number of vibrational states show that this error is a linear function of the LD parameter $\eta$.
To achieve the smallest error, one needs to correct the resonance position. We may now apply the formalism described in Section \ref{2_level_harmonic_section} to the Stark-Shift gate Hamiltonian (\ref{SS_ham_splitted}), 
with $\xi=\Omega_R$,  
see Appendix \ref{VBS_SS_gate_appendix}, to find that the positions of the structural and dynamical resonances for the $|+,n\ra\leftrightarrow|-,n+1\ra$ transition are shifted from the bare position $\Omega_R^{(0)}=\omega_T$ according to
\begin{eqnarray}
 \left(\Omega_R\right)_{S}&=&\omega_T-\frac{1}{4}\eta^2\omega_T(n+1),\\
\left(\Omega_R\right)_{D}&=&\omega_T+\frac{3}{4}\eta^2\omega_T(n+1).
\end{eqnarray}
%
Tuning the Rabi frequency to $(\Omega_R)_{D}$ does not change the dependance of the error with $\eta$, which remains linear, but 
improves by a factor of $\sim 3$ the obtained error, see Fig. \ref{error_eta_fig}a, while tuning to $(\Omega_R)_{S}$ the error is worse than for the bare resonance.
 
Since the Rabi frequency, and thus the gate velocity, are limited by the gate error \cite{steane00},
the threshold value $\epsilon_t$ of the imprecision of the gate 
will limit the processor speed. For a given required gate precision, 
the velocity of the SS gate (expressed as the inverse of the $\pi$-pulse duration) 
will be limited by the condition
\begin{equation}
\label{gate_vel_DR}
\frac{1}{T_n}=\frac{\eta\Omega_R\sqrt{n+1}}{\pi}\le\frac{\epsilon_t\omega_T}{\pi}.
\end{equation}
This bound can be improved by a factor of $\sim 3$ by optimization of the intensity as discussed above. 
%
The speed bound is better by $1/\eta$ than the bound of the gate of Monroe et al. \cite{Monroe} or the corrected CZ gate \cite{steane00}.

%


\subsection{Comparison with the CZ gate}
%
\begin{figure}[t]
\begin{center}
\includegraphics[width=7cm]{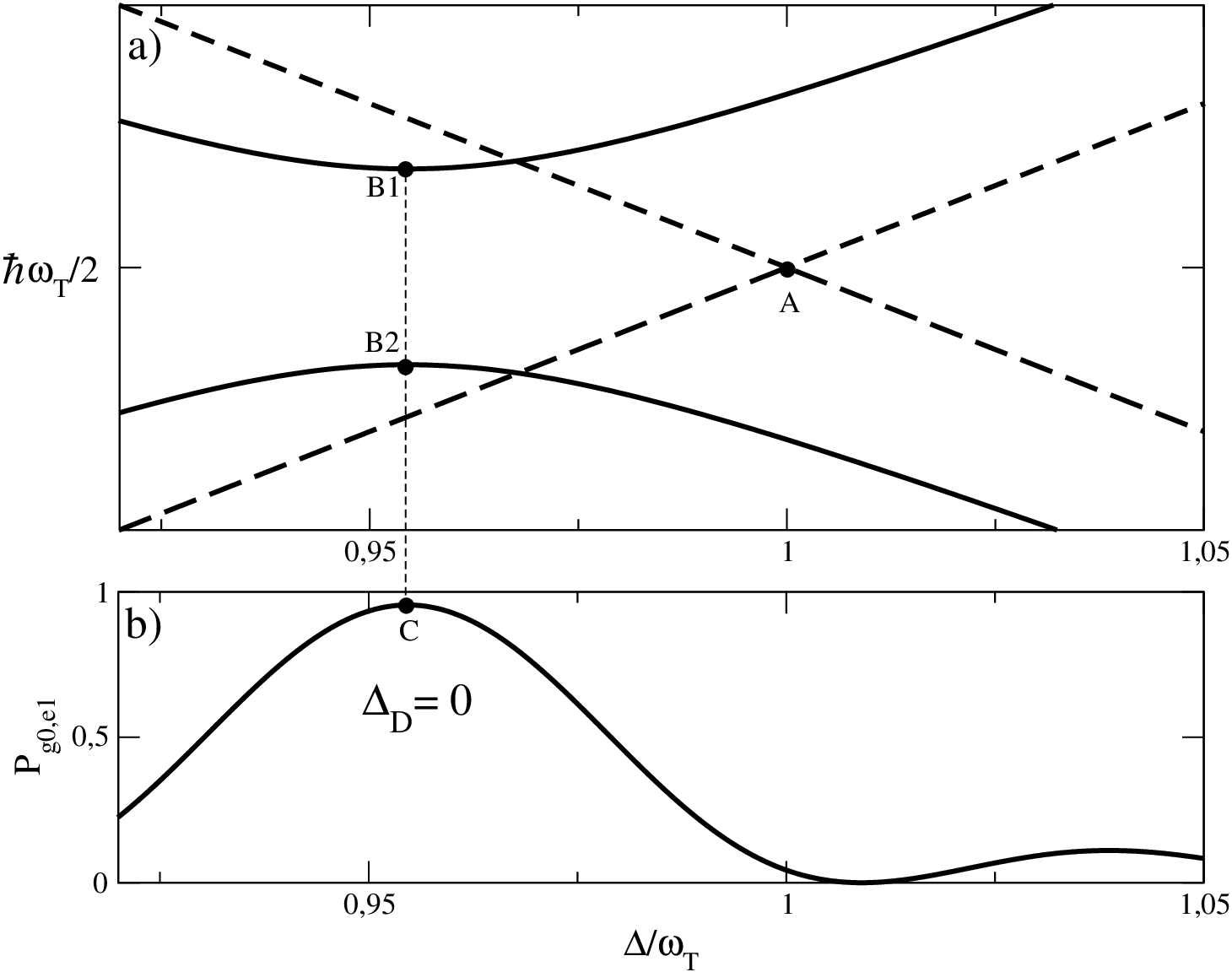}
\caption{Cirac-Zoller interaction (a) First few energy levels of a trapped ion at the first blue sideband, where the states $|g,0\ra$ and $|e,1\ra$ are on resonance.  Point $A$ marks the bare nominal resonance at $\Delta=\omega_T$ when the laser is 
turned off ($\Omega_R=0$), while points $B1$ and $B2$ define the structural resonance (the position of the anti-crossing) if the laser is turned on ($\Omega_R/\omega_T=0.3$). (b) State-flip probability from $|g,0\ra$ to $|e,1\ra$ after a $\pi$-pulse (defined by $\eta\Omega_R t=\pi$) has been applied; the maximum at $C$ defines the resonance location dynamically. 
In the CZ interaction both structural dynamical definitions coincide, there is no dynamical shift, i. e., $\Delta_D=0$.
All calculations have been done by numerically diagonalizing the full Hamiltonian (\ref{full_ham}) to all orders in $\eta$ and including a large number of vibrational states ($n=20$).}
\label{resonances_cz_fig}
\end{center}
\end{figure}

It is interesting to compare the SS scheme to the CZ gate scheme, where the perturbative parameter is no longer the LD parameter. In this case, the small parameter is the Rabi frequency (low intensity lasers), ``small'' meaning $\Omega_R\ll\omega_T$. Moreover, the computational basis in this scheme is the bare $|g\ra,|e\ra$ basis (for low intensity lasers the dressed states may be approximated by the bare states). The Hamiltonian (\ref{full_ham}) may then be partitioned as 
\begin{eqnarray}
\label{CZ_division}
H_0&=&\hbar\omega_T a^\dag a -\frac{\hbar\Delta}{2}\sigma_z,\nonumber\\
	V&=&\frac{\hbar\Omega_R}{2}\left[e^{i\eta(a+a^\dag)}\sigma_++H.c\right].
\end{eqnarray}
Applying once again the effective Hamiltonian formalism described in Section \ref{2_level_harmonic_section} (redefine $|a\ra=|g\ra$, $|b\ra=|e\ra$, and $\xi=\Delta$),
it is found that both structural and dynamical resonance definitions coincide, see Fig. \ref{resonances_cz_fig}. Thus, the CZ interaction shows no dynamical shift, $\Delta_D=0$.  In order to speed up the CZ gate, correcting for the structural shift is enough \cite{steane00}, see Fig. \ref{error_eta_fig}b.
Explicit expressions for the structural shift for a CZ type of interaction (the so-called Vibrational BS shift) have been provided elsewhere \cite{LME08}. 

\section{Discussion}
\label{ss_gate_discussion}

Two different definitions of a quantum resonance associated with an avoided crossing have been provided, which do not coincide in the general case. As an application, we have shown that, for the  
same precision requirements, the speed of the SS gate is of order $1/\eta$ times larger
than the speed of the corrected CZ gate, but its optimization implies
the new concept of 
tuning the laser intensity to the combined effect of dynamical and structural shifts.
We expect that this finding will have repercussions in other gates affected by light shifts \cite{SM,Leibfried}, and in related physical systems (cavity QED, superconducting qubits \cite{Deppe08}), in which a similar Hamiltonian structure and fast 
high intensity transitions are considered \cite{Solano}. The distinction between dynamical and structural shifts will also be relevant for metrological applications \cite{LME07}, as in atomic clocks and other interferometers. 

{\it Acknowledgments.} We acknowledge stimulating
discussions with C. Cohen-Tannoudji, and  
support by Ministerio de Educaci\'on y Ciencia (FIS2006-10268-C03-01) and the Basque Country University (GIU07/40).

\appendix
\section{Structural and Dynamical shifts for the SS gate interaction}
\label{VBS_SS_gate_appendix}
%
The matrix elements of the level shift operator $R$ and thus the structural or vibrational BS shift  for the CZ type of interaction (low intensity lasers) are provided in \cite{LME08}.
In order to obtain explicit expressions for the dynamical and structural resonances defined in Sec. \ref{2_level_harmonic_section} for the SS gate Hamiltonian, we calculate the matrix elements of the level shift operator using the SS-gate Hamiltonian given in Eq.  (\ref{SS_ham_splitted}), 
\begin{eqnarray}
r_{++}(E,\Omega_R)&=&\la +,n_+|r|+,n_+\ra=\frac{1}{2}\left(\Omega_{n_+,n_+}-\Omega_R\right)\nonumber\\
	&+&\frac{\hbar}{4}
\left[\sum_{k\neq n_+}\frac{\mathcal C_{k,n_+}^2}{E-\epsilon_{+,k}}
+\sum_{k\neq n_-}\frac{\mathcal{S}_{k,n_+}^2}{E-\epsilon_{-,k}}\right],\nonumber\\
r_{--}(E,\Omega_R)&=&\la -,n_-|r|-,n_-\ra=-\frac{1}{2}\left(\Omega_{n_-,n_-}-\Omega_R\right)\nonumber\\
&+&\frac{\hbar}{4}
\left[\sum_{k\neq n_-}\frac{\mathcal C_{k,n_-}^2}{E-\epsilon_{-,k}}
+\sum_{k\neq n_+}\frac{\mathcal{S}_{k,n_-}^2}{E-\epsilon_{+,k}}\right],\nonumber\\
r_{+-}(E,\Omega_R)&=&\la +,n_+|r|-,n_-\ra=i\frac{\Omega_R}{2}\mathcal{S}_{n_+,n_-},
\end{eqnarray}
with $\mathcal{C}_{n,n'}=\Omega_R\la n|\cos{\hat\alpha}|n'\ra=\rm{Re}\left(\Omega_{n,n'}\right)$ and 
$\mathcal{S}_{n,n'}=\Omega_R\la n|\sin\hat\alpha|n'\ra={\rm{Im}}\left(\Omega_{n,n'}\right)$ being the real and imaginary part of the coupling strenghts $\Omega_{n,n'}= \la n|e^{i\eta(a+a^\dag)}|n'\ra$ \cite{wineland98}, and where the non-perturbd energy levels are given by 
\begin{equation}
 \epsilon_{\pm,n}=n\hbar\omega_T\pm\frac{\hbar\Omega_R}{2}.
\end{equation}
%
As pointed out in \cite{LM07}, Rabi resonances will be well defined when the corresponding avoided crossings are well-isolated, 
i. e., when the ion is confined within the LD regime $\eta\ll1$ \cite{LM07}. In that 
case the matrix elements of $R$ to leading order in $\eta$ will only have contributions from the ``nearest neighbour'' vibrational states. The matrix elements are thus simplified to
 \begin{eqnarray}
r_{\pm,\pm}^{LD}&=&\eta^2\Omega_R\mathcal{F}_{\pm}(E,\Omega_R),\\
r_{+-}^{LD}&=&i\eta\frac{\Omega_R}{2}\sqrt {n_-} \delta_{n_+,n_--1},
 \end{eqnarray}
with
\begin{eqnarray} \mathcal{F}_\pm(E,\Omega_R)&=&\mp\frac{1}{2}\!\left(n_\pm+\frac{1}{2}\right)\nonumber\\
&\!+\!&\!\frac{\hbar\Omega_R}{4}\!\!\left(\!\!\frac{n_\pm}{E\!-\!\epsilon_{\mp,n_\pm-1}}\!+\!\frac{n_\pm+1}{E\!-\!\epsilon_{\mp,n_\pm+1}}\!\!\right)\!,
\end{eqnarray}
and where any term with a zero in the denominator has to be omitted since it is excluded from the sum.
Since these matrix elements are part of the implicit Hamiltonian (\ref{Heff}), the first approximation is evaluated at the non-perturbed position of the resonance, i. e., at $E=E_0=(n_++n_-)\hbar\omega_T/2$ and $\Omega_R=\Omega_R^{(0)}=(n_--n_+)\omega_T$.

With these expressions it is easy to find the structural shift for an arbitrary resonance, $\Delta_{S}=r_{--}-r_{++}+\Delta_D$, see Eq. (\ref{SR_2level_harmonic}). In particular, we find for the first few resonances
\begin{eqnarray}
(\Delta_{S})_{n,n+1}&=&-\frac{1}{4}\left(n+1\right)\eta^2\omega_T,\nonumber\\
(\Delta_{S})_{n,n+2}&=&-\frac{1}{3}\left(2n+3\right)\eta^2\omega_T,\nonumber\\
(\Delta_{S})_{n,n+3}&=&-\frac{3}{8}\left(n+2\right)\eta^2\omega_T,\nonumber\\
(\Delta_{S})_{n,n+4}&=&-\frac{2}{15}\left(2n+5\right)\eta^2\omega_T.
\end{eqnarray}
For the particular case of the SS-gate transition $|+,n\ra\leftrightarrow|-,n+1\ra$ discussed in Sec. \ref{ss_tuning_sec}, we have, to leading order in $\eta$,
\begin{eqnarray}
 r_{++}(\Omega_R=\omega_T)&=&-\frac{1}{8}\eta^2\omega_T(3n+2),\nonumber\\
r_{--}(\Omega_R=\omega_T)&=&\frac{1}{8}\eta^2\omega_T(3n+4),
\nonumber\\
\left.\frac{\partial\left|r_{+-}\right|^2}{\partial\Omega_R}\right|_{\Omega_R=\omega_T}&=&\eta^2\frac{\omega_T}{2}(n+1).
\label{R_elements_sstran}
\end{eqnarray}
The positions of the resonance according to structural and dynamical 
definitions for this SS-gate transition (as well as the value of the dynamical shift) are now readily calculated using Eqs. (\ref{SR_2level_harmonic}) and (\ref{DR_2level_harmonic}),
\begin{eqnarray} (\Omega_R)_{S}&=&\Omega_R^{(0)}+r_{--}-r_{++}-2\frac{\partial\left|r_{+-}\right|^2}{\partial\Omega_R}\nonumber\\
&=&\omega_T-\frac{1}{4}\eta^2\omega_T(n+1),\\
 (\Omega_R)_{D}&=&\Omega_R^{(0)}+r_{--}-r_{++}\nonumber\\
&=&\omega_T+\frac{3}{4}\eta^2\omega_T(n+1),
\end{eqnarray}
which corresponds to a dynamcial shift $|\Delta_D|=\eta^2\omega_T(n+1)$. Note that the derivatives of the diagonal terms $r_{--}$ and $r_{--}$ for computing the dynamical shift from Eq. (\ref{2_level_dyn_shift}) have been neglected, since they only contribute with $\eta^4$ terms.


\end{document}